\documentclass[%
 reprint,
 amsmath,amssymb,
 aps,
]{revtex4-1}

\usepackage{graphicx}
\usepackage{dcolumn}
\usepackage{bm}
\usepackage{epstopdf}
\usepackage{float}
\usepackage{amssymb}
\usepackage{amsfonts}
\usepackage{amsmath}

\begin{document}

\preprint{XXXX}

\title{Robust Scaling of Strength and Elastic Constants and Universal Cooperativity in Disordered Colloidal Micropillars}

\author{Daniel J. Strickland}
 \affiliation{Department of Materials Science and Engineering, University of Pennsylvania}
\author{Yun-Ru Huang}%
 \affiliation{Department of Chemical and Biological Engineering, University of Pennsylvania}
\author{Daeyeon Lee}%
 \affiliation{Department of Chemical and Biological Engineering, University of Pennsylvania}
\author{Daniel S. Gianola}%
 \affiliation{Department of Materials Science and Engineering, University of Pennsylvania} 

\begin{abstract}
{We study the uniaxial compressive behavior of disordered colloidal free-standing micropillars composed of a bidisperse mixture of 3 and 6 \textbf{$\mu m$} polystyrene particles.  Mechanical annealing of confined pillars enables variation of the packing fraction across the phase space of colloidal glasses. The measured normalized strengths and elastic moduli of the annealed freestanding micropillars span almost three orders-of-magnitude despite similar plastic morphology governed by shear banding.  We measure a robust correlation between ultimate strengths and elastic constants that is invariant to relative humidity, implying a critical strain of $\sim$0.01 that is strikingly similar to that observed in metallic glasses (MGs) [W.L. Johnson, K. Samwer, \textit{Phys. Rev. Lett}. \textbf{95}, 195501 (2005)] and suggestive of a universal mode of cooperative plastic deformation. We estimate the characteristic strain of the underlying cooperative plastic event by considering the energy necessary to create an Eshelby-like ellipsoidal inclusion in an elastic matrix.  We find that the characteristic strain is similar to that found in experiments and simulations of other disordered solids with distinct bonding and particle sizes, suggesting a universal criterion for the elastic to plastic transition in glassy materials with the capacity for finite plastic flow.}
\end{abstract}

\maketitle

In an ideal, defect-free system, the relationship between a material\textquoteright s elastic constants and yield strength is indicative of the underlying plastic event that generates macroscopic yielding.  For the case of a crystalline solid, Frenkel predicted the ideal yield stress by estimating the energy necessary to cooperatively shear pristine crystallographic planes. The result estimates that the ideal shear strength, $\tau_{y,ideal}$, scales linearly with the shear modulus, $\mu$, as    $\tau_{y,ideal} = \mu/2\pi$ \cite{Frenkel1926}.  This simple, yet striking, prediction implies a singular critical shear strain for yielding, irrespective of the material.  Experimental strengths of crystalline metals, however, are found to fall orders-of-magnitude short of Frenkel\textquoteright s ideal strength, suggestive of plasticity generated by mechanisms other than cooperative slip.  In real bulk crystals, microstructural defects, such as dislocations, grain boundaries, and surfaces, become plastically active at stresses well below $\tau_{y,ideal}$, implying a transition to a far less cooperative plastic deformation mechanism.  Controlling the character and number density of these defects mediates the strength of crystals; for instance, the Hall-Petch relationship for polycrystals \cite{0370-1301-64-9-303} \cite{Petch1953} states that $\tau_{y} \sim d^{-1/2}$ where $\tau_{y}$ is the yield strength and $d$ is the grain size (thus controlling the fraction of planar defects), and Taylor strengthening \cite{Taylor02071934} predicts $\tau_{y} \sim \rho^{1/2}$  where $\rho$  is the dislocation density. This ability to tailor material strength is a reflection of the large catalog of plastic events found in crystals and the associated broad range of energies necessary for their operation.  In such defected crystals, the highly cooperative shear mechanism that defines the intrinsic ideal strength is superseded by mechanisms that require the motion of only a few atoms (e.g. dislocation glide or climb), rather than the coordinated motion of many atoms. 

\begin{figure}[b]
	\centering
    \includegraphics[width=240 pt]{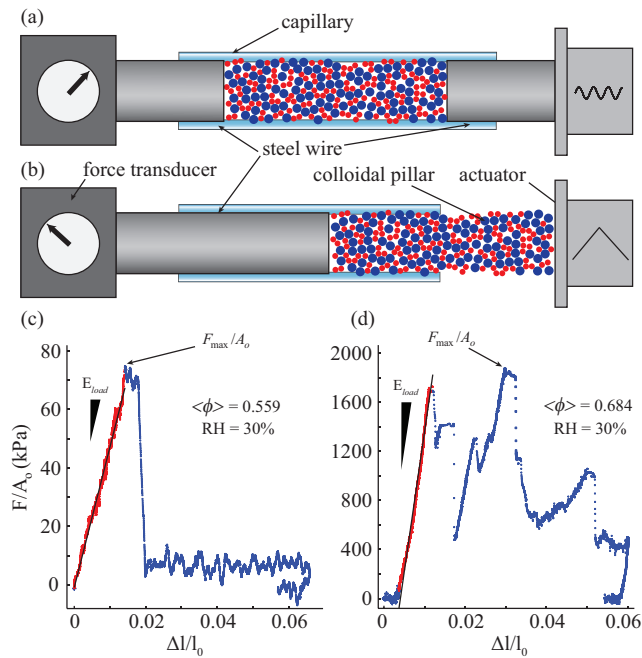}
    \caption{(a) The mechanical annealing setup.  The colloidal pillar is confined within a capillary by two steel wires between the actuator and the force transducer, and a sinusoidal displacement is generated by the actuator, leading to compaction of the pillar. (b) The uniaxial compression setup.  Extruding the pillar results in a freestanding sample that is compressed in situ in a confocal microscope. (c) Mechanical response of a lightly annealed pillar with $\langle\phi\rangle$ = 0.559 compressed at 30\% RH.  The measured effective elastic modulus is $E_{load}$ = 4.81 MPa and $\sigma_{max}$ = 74.8 kPa. (d) Mechanical response of a highly annealed pillar with  $\langle\phi\rangle$ = 0.684 compressed at 30\% RH, resulting in $E_{load}$ = 229 MPa and $\sigma_{max}$ = 1880 kPa. \label{Figure1}}
    
\end{figure} 

Metallic glasses (MGs) - amorphous alloys - on the other hand, exhibit a surprisingly robust relationship between yield strength and elastic constants despite their atomic heterogeneity and absence of long-range order.  Various MG alloys have been synthesized with shear moduli that range from 10 GPa to 80 GPa \cite{Inoue2011}.  Remarkably, an approximately universal elastic shear strain limit of $\gamma_{y} \approx 0.02$, where $\gamma_{y} = \tau_{y}/\mu$, has been measured \cite{Johnson2005} and directly demonstrated via atomistic simulations of MGs \cite{Cheng2011}.  Whereas structural modifications of the MGs through processes such as thermal treatment \cite{Wu2007}, severe plastic deformation \cite{Scudino2011}, and ion-beam irradiation \cite{Magagnosc2014}, \cite{Xiao2013} have been shown to alter the strength and ductility of MG alloys, the extent of strength variation is far below that of crystalline metals, which is suggestive of a single fundamental plastic event unique to MGs.  Furthermore, the characteristic failure mode observed in MGs -- \textquotedblleft shear banding\textquotedblright, in which plastic strain is localized in thin bands of the material --– is also found in amorphous solids composed of nanoparticles, colloids, and grains \cite{Desrues2002}, \cite{Chikkadi2011}, \cite{Zhang2013}.  The fact that a common failure mode is observed in amorphous solids with very different characteristic length scales and inter-particle interactions has led to the proposal that the fundamental plastic event found in MGs may in fact be universal to all amorphous solids.  Indeed, direct visualizations of sheared 3D colloid systems suggest cooperative shearing of collections of particles \cite{Chikkadi2011}, \cite{Schall2007}, \cite{Chikkadi2012}, although the link between these inelastic  building blocks and macroscopic yielding is still not clear.

In this article, we report on free-standing amorphous colloidal micropillars with compressive strengths that also exhibit a robust correlation with elastic constants and thus a universal elastic limit. By varying the packing fraction, $\phi$, we are able to vary the maximum transmitted force, equivalent to strength, and the elastic constants over almost three orders-of-magnitude.  We reconcile the robustness of the measured relationship by considering the energetics of the fundamental plastic event at criticality that is believed to underlie yielding in amorphous solids. This approach is based on the cooperative rearrangement first proposed by Argon following observations of sheared amorphous bubble rafts \cite{Argon1979}.  The idea has since been extended in several models, including the cooperative shear model (CSM) of Johnson and Samwer \cite{Johnson2005}, \cite{PhysRevLett.97.065502} and the shear transformation zone (STZ) theory of Falk and Langer \cite{PhysRevE.57.7192}, \cite{Langer2006}.  Our analysis results in an estimation of the characteristic transformation strain of a cooperative rearrangement with a magnitude that bears striking resemblance to that estimated in MGs, supporting the notion of a characteristic cooperative mechanism for plasticity in amorphous solids.  

We previously reported on a synthesis route for producing free-standing colloidal micropillars with cohesive particle-particle interactions \cite{C3CP55422H}.  Briefly, capillaries are filled with colloidal suspensions, subsequently dried, and carefully extruded to produce free-standing specimens for uniaxial mechanical testing.  The pillars relevant to the current work are 580 $\mu m$ in diameter and composed of a bidisperse mixture of 3.00 and 6.15 $\mu m$ diameter polystyrene (PS) spheres.  The mixture is prepared with a volume ratio $V_{L}/V_{S} = 3.78$, where $V_{L}$ and $V_{S}$ are the total volumes of the 6.15 $\mu m$ and 3.00 $\mu m$ spheres, respectively.  The bidisperse mixture is chosen to suppress crystallization; optical and electron microscopies are used to confirm amorphous packing.  

We developed a novel mechanical annealing procedure to alter the packing fraction $\phi$  -- and consequently the mechanical response -- which is described as follows.  After allowing the suspension of colloidal particles to dry within the capillary tube, two steel wires with diameters slightly smaller than the capillary diameter are inserted into both ends of the tube, rendering the packing fully confined (Figure ~\ref{Figure1}a).  A piezoelectric actuator is brought into contact with one of the wires and the opposite wire is coupled to a force transducer, enabling measurement of the axial force.  Sinusoidal displacements ($f = 0.5-5 \  Hz, A = 0.60-3.60 \  \mu m$) are produced by the actuator, which remains in contact with the wire.  The displacement periodically loads and unloads the pillar about the mean confining force ($F_{conf}  \sim 0.1-10 \ N$), resulting in a gradual densification of the pillar and an increase in $\phi$.  After mechanically annealing the pillar from one side, the capillary tube orientation is reversed and the process is repeated from the other side to promote uniform compaction. Varying the confining force in addition to the amplitude, frequency, and number of displacement cycles allows for some control of compaction.  Following annealing, the average packing fraction of the confined micropillar, $\langle\phi\rangle$, is determined by measuring the diameter, $D$, length, $L$, and mass, $m_{filled}$, using high-resolution optical microscopy and microbalance measurements, respectively.  Upon completion of mechanical testing, the mass of the empty capillary tube, $m_{empty}$, is measured.  Using the density of PS, $\rho_{PS}$, the packing fraction is determined as: \[\langle\phi\rangle = V_{solid}/V_{bulk} = \dfrac{(m_{filled}-m_{empty})/\rho_{PS}}{\pi D^2L/4}\] Our mechanical annealing procedure produced pillars with $0.528 \leq \langle\phi\rangle \leq 0.684$.  For our mixture of particles, the random close packing limit (RCP) is predicted to be $\phi  = 0.683$ \cite{Farr2009}.  The lower bound, random loose packing (RLP), for a bidisperse mixture of frictional, cohesive particles is not known, but rheology measurements on a bidisperse mixture of hard-spheres with similar diameter ($\frac{D_L}{D_S}$) and total volume ($\frac{N_L V_L}{N_S V_S}$) ratios show a fluidity limit, marked by a large increase in viscosity, at $\phi  = 0.550$ \cite{Shapiro1992}. Thus, our pillars span the full spectrum of glass packing.   

\begin{figure*}[htpb]
	\centering
    \includegraphics[width=480 pt]{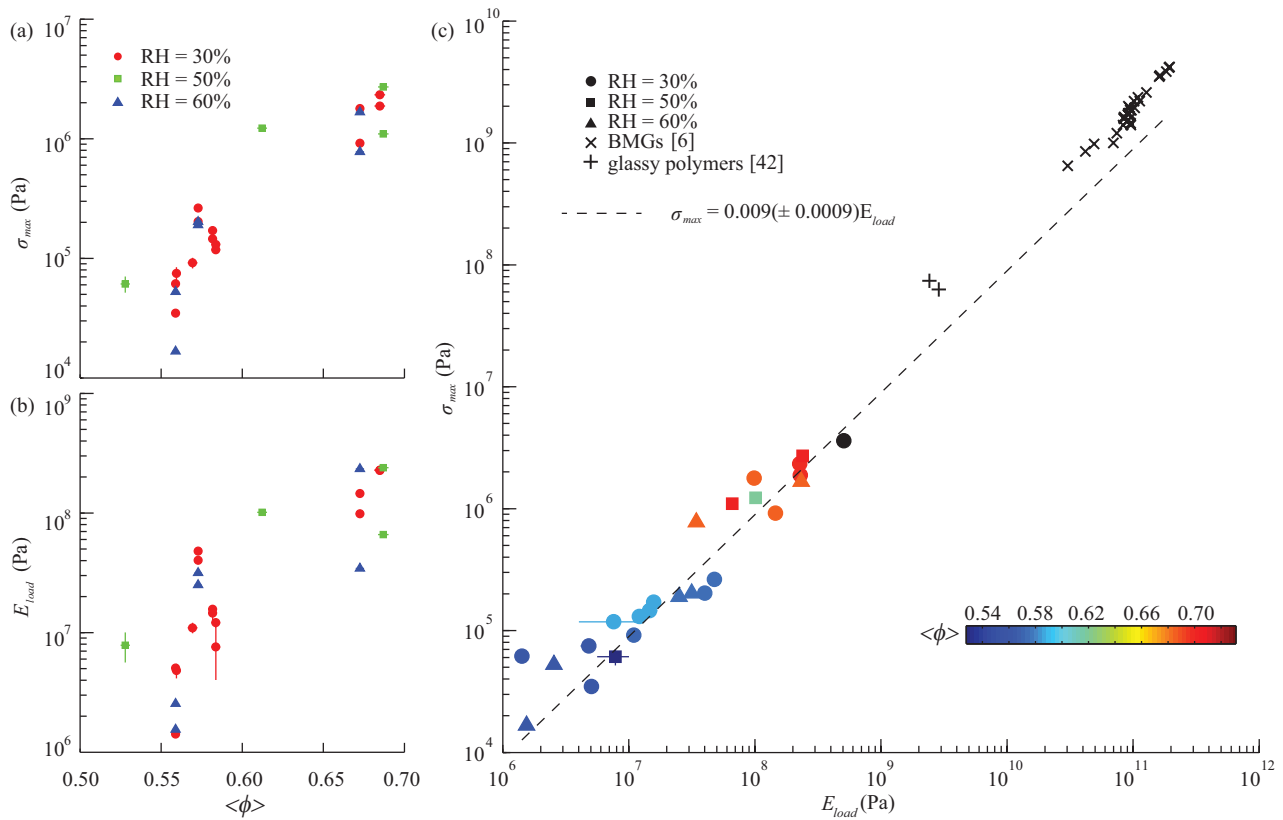}
    \caption{(a) The normalized maximum transmitted force, $\sigma_{max}$ vs. $\langle\phi\rangle$, (b) the normalized stiffness, $E_{load}$, vs. $\langle\phi\rangle$.  See supplemental material for error analysis.  The error is smaller than the data marker for the majority of measurements. (c) The normalized maximum transmitted force, $\sigma_{max}$, vs. $E_{load}$ showing a robust correlation that is relatively invariant with $\langle\phi\rangle$ (represented by data marker color;  black represents pillars where $\langle\phi\rangle$  measurements were not available).  Colloidal micropillar measurements are compared with yield strength and Young’s modulus values for metallic glasses \cite{Johnson2005} and glassy polymers \cite{Kozey1994}.  The dashed line shows the best linear fit to the colloidal data with slope $0.009\pm 0.0009$ representing the critical strain for failure.  See supplemental material for error analysis.  The error is smaller than the data marker for the majority of measurements. \label{Figure2}}
   
\end{figure*} 

\begin{figure}[b]
	\centering
    \includegraphics[width=240 pt]{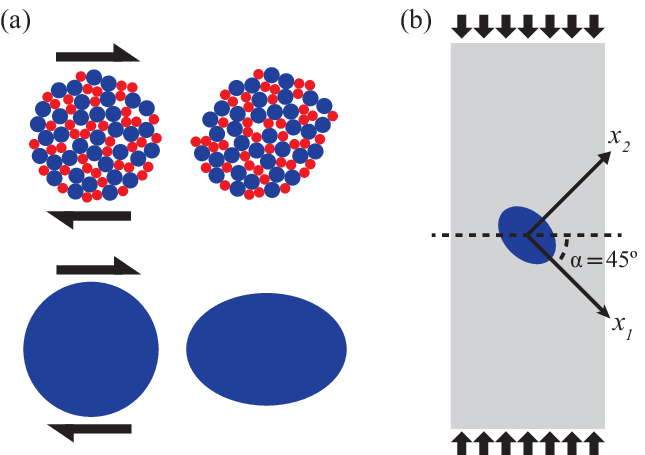}
    \caption{(a) Top, an idealized cooperative rearrangement induced by an applied shear stress.  Bottom, a continuum representation of the rearrangement. (b) The reference axes defined for the energy analysis.  The ellipsoid\textquoteright s major axis, $a$, lies along $x_1$ and its minor axis, $c$, lies along $x_2$. \label{Figure3}}
    
\end{figure}

Following our annealing steps, the micropillars were prepared for uniaxial compression.  The micropillars are made free-standing by extruding a desired length from the capillary using a precision drive-screw (Figure ~\ref{Figure1}b).  As the pillars are oriented with their major axes perpendicular to the force of gravity, the stability of our free-standing micropillars suggests strong cohesive particle-particle interactions. Between one and four free-standing specimens are obtained from each capillary and thus fluctuations in $\langle\phi\rangle$ from specimen to specimen are not quantified.  We opted for the pillar geometry owing to both a nominally simple (uniaxial) stress state and for facile comparison with previous work on MGs \cite{Schuh2007a}, \cite{Chen2008}, \cite{Greer2013a}.  Our experimental setup also contains an environmental chamber, enabling control of the relative humidity (RH) during testing.  Our previous work has shown that RH can alter the stiffness of a single pillar, which we attribute to variations in the amount of water contained within the pillar structure, affecting the number and connectivity of capillary bridges that form between particles (and thus the cohesive interactions) \cite{C3CP55422H}.  Reflectance confocal micrographs acquired during compression are used to digitally determine the displacement of the punch and the base of the pillar and therefore allow axial strain within the pillar to be calculated.  For more details on the experimental setup, refer to \cite{C3CP55422H}. 

Mechanical responses for specimens with   $\langle\phi\rangle = 0.559$ and $\langle\phi\rangle = 0.684$ are shown in Figure ~\ref{Figure1}c and ~\ref{Figure1}d.  Both specimens exhibit an initial loading regime where the transmitted force, $F$, increases linearly with $\Delta l/l_o$ ($\Delta l$ and $l_o$ are the change in length and undeformed length of the pillar, respectively).  For the $\langle\phi\rangle = 0.559$ specimen, significant structural evolution is observed in the micrographs for $\Delta l/l_o > \Delta l/l_o\vert_{F_{max}}$ and the subsequent load drop correlates with the development of a system-spanning shear band.  Further displacement shears the pillar along the band until final fracture. The $\langle\phi\rangle = 0.684$ specimen exhibits a series of elastic loadings and load drops that correspond with the formation of multiple shear bands.  A plastic morphology characterized by one or more shear bands, typically emanating from the punch, is consistently observed irrespective of $\langle\phi\rangle$.  We define the ultimate compressive strength of the micropillars as the maximum normalized force, $F_{max}/A_o \equiv \sigma_{max}$ ($A_o$ is the cross sectional area of the undeformed pillar), sustained by a pillar during a compression cycle.  We also define an effective elastic modulus upon loading, $E_{load}$, based on a linear fit to $F/A_o$ vs. $\Delta l/l_o$  between initial loading and the first yielding event.  Previously, we have measured the mechanical dissipation in a dense micropillar at different RHs \cite{C3CP55422H}.  Briefly, we found that at RHs up to $\sim$30\%, $\sim$80\% of the work done on loading was recovered on unloading for small total strains ($\Delta l/l_o \sim 0.003$).  Increasing RH beyond 30\% reduced the amount of energy recovered on unloading.  Neglecting dissipation results in an underestimation of the elastic component of stiffness (see analysis in Supporting Information); the consequences of underestimating the true elastic modulus will be discussed later.  

The measured ultimate strengths and effective elastic moduli for 27 micropillar specimens are shown in Figure ~\ref{Figure2}a and ~\ref{Figure2}b.  Both $F_{max}/A_o$ and $E_{load}$ are found to vary by more than 2.5 orders-of-magnitude over the range of $\langle\phi\rangle$ studied, corresponding to relatively loose and dense packings in the limit of low and high $\langle\phi\rangle$, respectively. We contend that $\sigma_{max}$ is the best measure of the intrinsic strength of the pillar because plastic activity localized near the punch, which occurs at lower levels of applied stress, is likely a result of surface roughness and not  representative of the of the bulk structure of the micropillar.  Surprisingly, $\sigma_{max}$ and $E_{load}$ appear to be relatively insensitive to RH in the range covered in these experiments. 

Our measurements demonstrate a strong correlation between $\sigma_{max}$ and $E_{load}$ of individual disordered micropillars (Figure ~\ref{Figure2}c), with a slope of $0.009\pm0.0009$ that is invariant to $\langle\phi\rangle$ and RH (shown by coloring and shape of markers, respectively, in Figure ~\ref{Figure2}c).  The robust relationship between $\sigma_{max}$ and $E_{load}$ is suggestive of a unique and cooperative plastic event that establishes the maximum strain the micropillars can withstand before macroscopically failing.  The insensitivity of this universal scaling to RH also suggests that the details of inter-particle interactions do not influence the critical strain, which implies a plasticity mechanism that is universal to disordered solids with the capacity for finite plastic flow.  This notion is further bolstered by the remarkable similarity of our measured scaling to that compiled from compressive behavior of MGs by Johnson and Samwer \cite{Johnson2005} as shown in Figure ~\ref{Figure2}c.  We further note that such scaling between strength and elastic constants has been reported in atomistic simulations of nanocrystalline alloys \cite{Rupert2014} which in the limit of diminishingly small grain sizes have been shown to exhibit cooperative mechanisms of plasticity reminiscent of metallic glasses (e.g. shear banding, pressure-sensitive yield criteria) \cite{Trelewicz2007}.

To understand the scaling relationship in our relatively athermal colloidal micropillars, we model the fundamental building block of cooperative plastic flow in the framework of Eshelby-like elasticity. Specifically, we consider the change in free energy associated with the introduction of an ellipsoidal inclusion -- representing the cooperative shear transformation -- in an elastic matrix subjected to an applied far-field stress.  This approach is motivated by experiments \cite{Argon1979}, \cite{Schall2007}, \cite{Keim2013} and simulations \cite{Deng1989}, \cite{doi:10.1080/01418618108239553} on the deformation of amorphous solids that suggest that the fundamental plastic event is a cooperative, shear-induced rearrangement of $\sim$10-100  particles \cite{Argon1979a} referred to as a shear transformation zone (STZ) (see Figure ~\ref{Figure3}).  After operation, in which the STZ evolves from the initial to the deformed state, a local elastic strain field is generated in the STZ and the surrounding matrix owing to elastic compatibility.  This elastic interaction has been analyzed using Eshelby\textquoteright s solution for the introduction of an inclusion in an elastic matrix \cite{Eshelby1957}, \cite{PhysRevE.88.032401}, \cite{Homer2014}.  In addition to the elastic energy of the confined shear transformation, the far-field stress deforms the inclusion, thereby resulting in work done.  Taken together, the elastic energy and work terms yield a simple expression for the change in free energy associated with the introduction of the inclusion:
\begin{equation}
\begin{split}
\Delta G &= -\dfrac{1}{2}\int_\Omega \! \sigma_{ij}^I \epsilon_{ij}^T \, \mathrm{d}V 
-\int_\Omega \! \sigma_{ij}^{\infty} \epsilon_{ij}^T \, \mathrm{d}V \\ &= -\dfrac{\Omega}{2}\sigma_{ij}^I \epsilon_{ij}^T - \Omega \sigma_{ij}^\infty \epsilon_{ij}^T \label{qg1}
\end{split}
\end{equation}
Here, $\sigma_{ij}^I$ is the stress field inside the confined inclusion, $\epsilon_{ij}^T$ is the unconfined transformation strain of the inclusion,  $\sigma_{ij}^{\infty}$ is the applied far-field stress, and $\Omega$ is the volume of the inclusion.  The integrals are readily evaluated because the stress and strain fields inside of an ellipsoidal inclusion are spatially uniform.  The stress field inside the inclusion can be written as:
\begin{equation}
\sigma_{ij}^I = C_{ijkl}(S_{klmn}-\delta_{km} \delta_{ln})\epsilon_{mn}^T  \label{qg2}
\end{equation}
Where $C_{ijkl}$ is the stiffness tensor, $S_{klmn}$ is Eshelby\textquoteright s tensor, and $\delta_{ij}$ is the Kronecker delta.  Eshelby\textquoteright s tensor relates the unconfined transformation strain of the inclusion to the confined strain of the inclusion (i.e., the strain after being reinserted in the matrix) \cite{Eshelby1957}. The shear bands that form in the micropillars are oriented approximately $45^{\circ}$ from the pillar axis, which is similar to the orientation of shear bands found in compressed BMGs \cite{Greer2013a} and soil pillars \cite{Desrues2002}. While it is believed that the nature of external loading may bias the orientation of shear bands  (towards the pillar axis in compression \cite{Gao2011}), we lack the ability to measure shear band orientation to such precision.  Therefore, for the energy analysis we neglect any strong pressure-dependent yielding and assume a tri-axial (i.e., each axis is unique in its length, $a>b>c$) ellipsoidal inclusion with the major axis, $a$, lying along the direction of maximum shear stress, $\alpha = 45^{\circ}$ (see Figure ~\ref{Figure3}).  We also assume that the middle axis, $b$, of the ellipsoid is neutral, in which case the transformation strain tensor reduces to the form defined by a basis set parallel to the ellipsoidal axes:
\[
\epsilon_{ij}^T  = \epsilon^\ast\begin{array}{lcl}\begin{bmatrix}
0 & 1 & 0 \\
1 & 0 & 0 \\
0 & 0 & 0
\end{bmatrix}
\end{array}
\]
which can be described by the scalar strain magnitude $\epsilon^{\ast}$.  Assuming an isotropic elastic medium, Equation 2 reduces to: 
\[
\sigma^I = \dfrac{E(7-5\nu)}{15(1-\nu^2)}\epsilon^{\ast}
\]
where $E$ is Young’s modulus and $\nu$  is Poisson’s ratio \cite{PhysRevE.88.032401}.  This expression represents the self-stress of the inclusion, which is completely defined by the material\textquoteright s elastic constants, $E$ and $\nu$, and $\epsilon^{\ast}$.  For isotropic compression and the reference basis defined in Figure ~\ref{Figure3}, $\sigma_{ij}^\infty$ can be written as:
\[
\sigma_{ij}^\infty  = -\dfrac{\sigma}{2}
\begin{array}{lcl}\begin{bmatrix}
1 & 1 & 0 \\
1 & 1 & 0 \\
0 & 0 & 0
\end{bmatrix}
\end{array}
\]
where $\sigma = F/A_o$ is the applied stress, and compression is negative.  Equation 1 then reduces to:
\begin{equation}
\Delta G = -\dfrac{\Omega}{2}\dfrac{E(7-5\nu)}{15(1-\nu^2)}{\epsilon^{\ast}}^2 + \Omega \sigma \epsilon^{\ast} \label{qg3}
\end{equation}
We assume that at $\sigma_{max}$, $\Delta G = 0$, and upon a further increase in the applied stress, the introduction of an inclusion (i.e. operation of a shear transformation) becomes energetically favorable.  The relationship between strength and stiffness thus becomes:
\begin{equation}
\sigma_{max} = \dfrac{E(7-5\nu)}{30(1-\nu^2)}\epsilon^{\ast} 
\label{qg4}
\end{equation}

We assume $E = E_{load}$ and values of $\nu$  between $0.15$ \cite{Zhang2013} and $0.45$ and find a best fit for the data with   $\epsilon^{\ast}$ as the free parameter.  Over the range of  considered, $\epsilon^{\ast}$ ranges from 0.042 for $\nu  = 0.15$ to $\epsilon^{\ast} = 0.045$ for  $\nu  = 0.45$; thus $\epsilon^{\ast}$ is largely insensitive to $\nu$. Because our system is dissipative, the true elastic modulus is larger than the stiffness measured on loading.  Assuming that 50\% of the work done on the system during loading is stored as elastic energy \cite{C3CP55422H}, the true elastic modulus is underestimated by a factor of 2 ($2E_{load}=E_{elastic}$, see Supporting Information).  This error results in an overestimation of $\epsilon^{\ast}$ by a factor of 2.  Therefore, we take $\frac{\epsilon^{\ast}}{2}$ and $\epsilon^{\ast}$ as bounds on the magnitude of the  characteristic transformation strain. The magnitude of $\epsilon^{\ast}$ found using this analysis of our micropillar data is similar to the values found in simulations of Lennard-Jones particles \cite{PhysRevE.88.032401} and experiments of sheared bubble rafts \cite{Argon1979} in which displacement fields can be measured directly.  Dasgupta et al. compared the non-affine displacement fields generated by an STZ in a molecular dynamics simulation to the displacement field generated by a general Eshelby transformation strain \cite{PhysRevE.88.032401}.  These authors found good agreement between the fields when using a traceless Eshelby transformation strain with two non-zero eigenvalues and $\epsilon^{\ast} = 0.04$, which agrees well with our value of $\sim 0.04$.  Argon estimated a shear strain value $\gamma_o = 2\epsilon^{\ast} = 0.125$  based on observations of bubble rafts \cite{Argon1979a}.  Recent kinetic Monte Carlo (kMC) simulations of MGs that have been successful in capturing shear band formation employ a characteristic STZ strain $\gamma_o = 0.10$ when determining the free-energy change associated with STZ operation \cite{Homer2014}, \cite{Homer2009}, \cite{Homer2010}.  This similiarity in shear transformation kinematics surprisingly extends to other classes of amorphous solids.  Simulations of sheared amorphous silicon - a network glass with strongly directional bonding - show a characteristic transformation strain of $\sim$0.015 \cite{doi:10.1080/14786430600596852}.  The authors of this study note that while the characteristic size of shear transformations appears to be bonding dependent ($\sim$1 nm in metallic glasses, $\sim$3 nm in amorphous silicon, and $\sim$10 nm in glassy polymers \cite{doi:10.1080/14786430600596852}), the transformation strain remains similar across systems.  Indeed, glassy polymers that are known to develop shear bands upon yielding, such as polymethyl methacrylate (PMMA) \cite{doi:10.1080/14786437008225837} and PS \cite{Argon1968}, show critical strains similar to those of MGs ($\gamma_{y} \approx 0.045$ and $\gamma_{y} \approx 0.033$ for PMMA and PS, respectively \cite{Kozey1994}). The robust critical strain appears to break down in glasses which show deformation morphology other than shear banding. For example, the critical strain in amorphous silica nanowires that exhibit brittle behavior and cleavage fracture is $\gamma_{y} \approx 0.2$ \cite{Brambilla2009}, much larger than the value found in MGs and glassy polymers. Thus, the cooperative shear mechanism discussed in this work hinges on the intrinsic capacity for plastic flow that precedes final fracture. 

This simple model does not capture the complex dynamical interaction of activated and nucleating STZs that determine the ultimate deformation morphology, which likely governs the extent of plastic deformation and the spatio-temporal evolution from individual STZ operation to macroscopic shear localization.  However, the robustness of the correlation between $\sigma_{max}$ and $E_{load}$ for a wide range of structural configurations brought about by mechanical annealing suggests that incipient operation of STZs and macroscopic plastic flow along shear planes occur nearly simultaneously.  In other words, the transition from the quasi-elastic to plastic regimes is sharp with respect to stress.  This can be inferred as a signature of a system driven in the athermal limit with a relatively narrow distribution of barrier energies defining the fundamental unit of plastic deformation.  In contrast to thermal systems, such as metallic glasses, maneuvering within the complex potential energy landscape of our athermal colloidal systems is not aided by thermal activation.  We assert that our system is athermal by considering the non-dimensional parameter $\frac{k_B T}{\epsilon}$, where $k_B$ is Boltzmann\textquoteright s constant, $T$ is the temperature, and $\epsilon$ is a measure of the interaction energy between particles assuming Hertzian contact.  This parameter is a measure of the thermal energy relative to the elastic energy stored in the particles and vanishes in the athermal limit. For our system, $\frac{k_B T}{\epsilon}\sim 1\times10^{-14}$, much less than the value found in other systems treated as athermal \cite{C3SM52454J}.  The significance of rate effects that could arise from capillary bridge formation is quantified in the parameter $\dot{\gamma} \tau$, where $\dot{\gamma}$ is the strain rate and $\tau$ is the timescale associated with the nucleation of water capillaries.  Assuming a capillary nucleation timescale similar to that measured on silicon surfaces \cite{Greiner2010}, $\dot{\gamma} \tau \sim 1\times10^{-11}$, indicating that nucleation events occur at timescales much smaller than the timescale associated with the imposed strain.  With thermal fluctuations absent, the applied stress alone surmounts the local energy maxima, ultimately driving the cooperative events.  In turn, the compatibility constraint of the elastic matrix upon shear transformation (cooperative rearrangement of a collection of particles with a characteristic strain $\epsilon^{\ast}$) provides the long-range interaction to drive localized failure.  Taken as a whole, the similarities in macroscopic yielding strain, characteristic STZ strain, and shear band morphology between our colloidal packings and metallic glasses, despite the dissipative nature of our particle-particle interactions, lend support to the notion of a universal, cooperative plastic event unique to amorphous solids with the capacity for plastic flow.

\begin{acknowledgments}
We gratefully acknowledge financial support from the National Science Foundation through PENN MRSEC DMR-1120901.  We thank the Penn Nanoscale Characterization Facility for technical support and D.J. Magagnosc for technical assistance and insightful discussions. We are also grateful to A. Liu for critical reading of our manuscript and insightful comments.
\end{acknowledgments}

\bibliographystyle{hieeetr}   
\bibliography{pnas2014}

\end{document}